# Competition of Chiroptical Effect Caused by Nanostructure and Chiral Molecules

*Tong Wu, Jun Ren, Rongyao Wang and Xiangdong Zhang\**

School of Physics, Beijing Institute of Technology, Beijing, 100081, China

**ABSTRACT:**   The theory to calculate circular dichroism (CD) of chiral molecules in a finite cluster with arbitrarily disposed objects has been developed by means of T-matrix method. The interactions between chiral molecules and nanostructures have been investigated. Our studies focus on the case of chiral molecules inserted into plasmonic hot spots of nanostructures. Our results show that the total CD of the system with two chiral molecules is not sum for two cases when two chiral molecules inserted respectively into the hot spots of nanoparticle clusters as the distances among nanoparticles are small, although the relationship is established at the case of large interparticle distances. The plasmonic CD arising from structure chirality of nanocomposites depends strongly on the relative positions and orientations of nanospheroids, and are much greater than that from molecule-induced chirality. However, the molecule-induced plasmonic CD effect from the molecule-NP nanocomposites with special chiral structures can be spectrally distinguishable from the structure chirality-based optical activity. Our results provide a new theoretical framework for understanding the two different aspects of plasmonic CD effect in molecule-NP nanocomposites, which would be helpful for the experimental design of novel biosensors to realize ultrasensitive probe of chiral information of molecules by plasmon-based nanotechnology.

# I. INTRODUCTION

Chirality plays a pivotal role in biochemistry and the evolution of life itself[1-2]. Many biologically active molecules are chiral, including the naturally occurring amino acids and sugars, detection and quantification of chiral enantiomers of these biomolecules are of considerable importance for biomedical diagnostics and pathogen analyses. Circular dichroism (CD) spectroscopy is one of the central methods to probe chiral nature of molecules through describing the difference in absorption of right- and left-handed circularly-polarized photons [3-4]. In general, molecular CD signal is typically weak, thus, chiral analyses by such a spectroscopic technique have usually been restricted to the analyst at a relatively high concentration [1-4]. This is frequently an impediment for a practical use because an ultrasensitive probe of tiny amounts of chiral substance is highly demanded for practical biosensing applications in biomedical and pharmaceutical fields. Recent investigations have shown that a weak molecular CD signal in the UV spectral region can be both enhanced and transferred to the visible/near-infrared region when chiral molecules are adsorbed at the surfaces of metallic nanoparticles (NPs) or in the nanogaps (i.e. hot spots) of particles' clusters [5-14]. Thus, plasmon-enhanced CD spectroscopy is expected to be a promising solution for overcoming the probe obstacle of tiny amounts of chiral molecules in conventional electronic CD measurements[15-19].

However, some plasmonic nanostructures themselves (called chiral nanostructures) possess intrinsic chirality [20-29], such as helically arranged metallic NPs, twisted metallic nanorods and so on [27-29]. In a real experimental system, structural chirality may exist in the molecule-NP nanocomposites. That is to say, the CD signal from molecule-NP nanocomposites includes two aspects of contribution, chiral molecules (molecule-induced plasmon chirality) and chiral nanostructures (structure chirality). Recent experimental studies suggested that structure chirality-based plasmonic CD activity displayed less dependence on the molecular chirality[30], whereas the molecule-induced plasmonic CD from nonchiral NP clusters possessed a strong correlation with the chiral nature of molecules located at the hot spots[31]. Thus, quantifying the individual contribution of the molecule-induced optical activity to the overall plasmonic CD signal from the nanocomposites is necessary for the probe of molecular chirality. [32-36] However, so far little is known whether the plasmonic CD effects arising from the above two different aspects are distinguishable in a molecule-NP nanocomposite.

In this work we present a method to calculate CD of molecule-NP nanocomposites by using the T-matrix[37-38]. Based on such a method, we study both the molecule-induced and structure chirality-based plasmonic CD effects occurring in some molecule-NP nanocomposites. We show that plasmonic CD arising from structure chirality in some nanocomposites depends strongly on the relative positions and orientations of nanospheroids, and are much greater than that from molecule-induced chirality. On the other hand, for the nanocomposites without structure chirality, strong near-field plasmonic coupling between molecules and NPs plays crucial role in determining the overall molecule-induced CD spectral responses. Furthermore, we also find that the molecule-induced plasmonic CD effect from the molecule-NP nanocomposites with special chiral structures can be spectrally distinguishable from the structure chirality-based optical activity.

**II.Theory and method.**

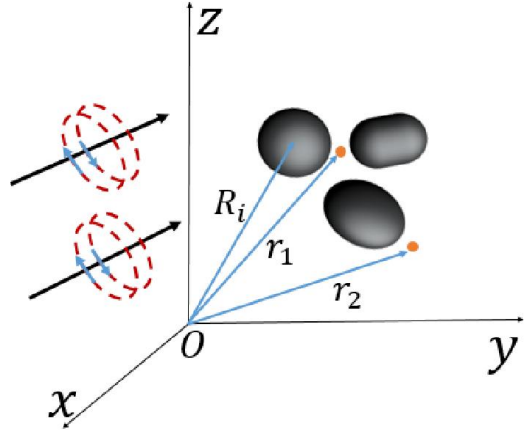

Figure 1. (Color online) Geometry and coordinate of a hybrid system consisting of two molecules and a finite cluster with arbitrarily disposed objects. The system is scattered by an incident polarized light, $\vec{r}_1$ and $\vec{r}_2$ are position vectors of molecule 1 and 2, $\boldsymbol{R}_i$ represents the position of the $i^{th}$ disposed object.

We consider here a semi-classical hybrid system consisting of two molecules and a cluster of n NPs as shown in Figure 1, which is excited by circularly polarized light $\vec{E}_0(\omega)$. The molecules used here are assumed to be point-like two-level systems, no vibrational structure of transitions is considered. The parameters of the two molecules are taken identical without possibility of inhomogeneous broadening. According to the previous investigation [5], the contribution of electric quadrupole terms to the total CD signal is very trivial because of the symmetry in some hybrid systems. Thus, they are excluded in the following discussions. A CD signal from the system is defined as the difference between absorption of left- and right-handed polarized lights, which can be written as [5, 39]

$$CD = \langle Q_+ - Q_- \rangle_\Omega, \qquad (1)$$

where the averaging over the solid angle, $\langle \cdots \rangle_\Omega$, is needed since hybrid nanoparticle complexes have random orientations. $\langle Q_+ \rangle$ and $\langle Q_- \rangle$ represent the absorption rate for the left- and right-handed polarized light, respectively. $Q_\pm = Q_{mol\pm} + Q_{NP\pm}$, where $Q_{mol\pm}$ and $Q_{NP\pm}$ are the absorption rate for the molecules and the nanoparticle cluster, respectively. In general, the total CD ($CD_{total}$) of the system can be divided into two parts.

$$CD_{total} = CD_{mol} + CD_{NP}, \qquad (2)$$

where $CD_{mol} = \langle Q_{mol+} - Q_{mol-} \rangle_\Omega$ and $CD_{NP} = \langle Q_{NP+} - Q_{NP-} \rangle_\Omega$. In the following we calculate the absorption rate and CD for two parts, respectively.

## A. Absorption rate of chiral molecules and $CD_{mol}$

For the above system, the master equation for quantum states of the $\kappa$ molecule can be written as

$$\hbar \frac{\partial \rho_{ij}^\kappa}{\partial t} = i \left[ \hat{\rho}^\kappa, \hat{H}^\kappa \right]_{ij} - \Gamma_{ij}^\kappa(\rho), \qquad (3)$$

where $\hat{H}^\kappa = \hat{H}_0^\kappa + H_1^\kappa$ is the Hamiltonian of the $\kappa$ molecule ($\kappa = 1, 2$), $\hat{\rho}^\kappa$ is the density matrix and $\rho_{21}^\kappa = \sigma_{21}^\kappa e^{-i\omega t}$ is the corresponding matrix element. Here $\hat{H}_0^\kappa$ describes the internal electronic structure of the $\kappa$ molecule and $\hat{H}_1^\kappa = -\hat{\boldsymbol{\mu}}^\kappa \cdot \boldsymbol{E}_T - \hat{\boldsymbol{m}}^\kappa \cdot \boldsymbol{B}_T$ is the light-matter interaction operator, $\boldsymbol{E}_T = \boldsymbol{E}_{tot} e^{-i\omega t} + \boldsymbol{E}_{tot}^* e^{i\omega t}$ and $\boldsymbol{B}_T = \boldsymbol{B}_{tot} e^{-i\omega t} + \boldsymbol{B}_{tot}^* e^{i\omega t}$ are the total fields acting on the molecule. Also $\hat{\boldsymbol{\mu}}^\kappa = e \boldsymbol{r}^\kappa$ and $\hat{\boldsymbol{m}}^\kappa = \dfrac{e}{2m_e} \boldsymbol{r}^\kappa \times \boldsymbol{p}^\kappa$ are electric and magnetic operators. $\Gamma_{ij}^\kappa(\rho)$ is the relaxation term which describes the phenomenological damping, and for only two quantum states, it is given by $\Gamma_{11}^\kappa(\rho^\kappa) = -\Gamma_{22}^\kappa(\rho^\kappa) = -\gamma_{22}\rho_{22}^\kappa$, $\Gamma_{12}^\kappa(\rho) = \gamma_{21}\rho_{12}^\kappa$ and $\Gamma_{21}^\kappa(\rho) = \gamma_{21}\rho_{21}^\kappa$.

Specifically the total electric field satisfies $\boldsymbol{E}_{tot} = \boldsymbol{E}' + \boldsymbol{E}_{ds}^\kappa + \boldsymbol{E}_{ds}^{\kappa'}$ and $\boldsymbol{E}' = \boldsymbol{E}_0 + \boldsymbol{E}_s$. Here $\boldsymbol{E}_0$ is the electric field of the incident wave and $\boldsymbol{E}_s$ is the field induced by NPs in the absence of molecules, $\boldsymbol{E}_{ds}^\kappa$ and $\boldsymbol{E}_{ds}^{\kappa'}$ are respectively the scattered field of molecule $\kappa$ and $\kappa'$ by NPs ($\kappa \neq \kappa'$). The molecule $\kappa$ has a dipole moment of $\boldsymbol{d}_{mol}^\kappa = \text{Tr}(\hat{\rho}^\kappa \hat{\boldsymbol{\mu}}^\kappa) = \boldsymbol{d}_\kappa e^{-i\omega t} + \boldsymbol{d}_\kappa^* e^{i\omega t}$ with $\boldsymbol{d}_\kappa = \sigma_{21}^\kappa \boldsymbol{\mu}_{12}^\kappa$ which are induced by the electromagnetic (EM) field. We have

$$\boldsymbol{E}_{ds}^\kappa = \sigma_{21}^\kappa \boldsymbol{E}_d^\kappa, \tag{4}$$

where $\boldsymbol{E}_d^\kappa$ can be viewed as the complex electric field generated by a dipole with moment $\boldsymbol{\mu}_{12}^\kappa$ located in $\boldsymbol{r}_\kappa$. Here we assume that the molecules are sufficiently far from each other, so that intermolecular Coulomb interactions can be ignored. Considering that the magnetic field in a small structure is not affected much by plasmonic effects, the approximation $\boldsymbol{B}_{tot} \approx \boldsymbol{B}' \approx \boldsymbol{B}_0$ is taken. Using the rotating wave approximation and treating the molecule as two level energy systems, we have

$$\hbar \frac{\partial \sigma_{21}^\kappa}{\partial t} \approx \left[i\hbar(\omega - \omega_0) - \gamma_{21}\right]\sigma_{21}^\kappa + i\left(\rho_{11}^\kappa - \rho_{22}^\kappa\right)\left(\boldsymbol{\mu}_{21}^\kappa \cdot \boldsymbol{E}_{tot}^\kappa + \boldsymbol{m}_{21}^\kappa \cdot \boldsymbol{B}_{tot}^\kappa\right), \tag{5}$$

$$\hbar \frac{\partial \rho_{22}^\kappa}{\partial t} \approx -i\left[\sigma_{21}^\kappa\left(\boldsymbol{\mu}_{12}^\kappa \cdot \boldsymbol{E}_{tot}^{\kappa*} + \boldsymbol{m}_{12}^\kappa \cdot \boldsymbol{B}_{tot}^{\kappa*}\right) - \sigma_{12}^\kappa\left(\boldsymbol{\mu}_{21}^\kappa \cdot \boldsymbol{E}_{tot}^\kappa + \boldsymbol{m}_{21}^\kappa \cdot \boldsymbol{B}_{tot}^\kappa\right)\right] - \gamma_{22}\rho_{22}^\kappa. \tag{6}$$

Here $\omega_0 = \omega_2 - \omega_1$ is the frequency of molecular transition. Considering $\sigma_{21}^\kappa$ and $\rho_{22}^\kappa$ are slowly varying variable, thus we have $\hbar \frac{\partial \sigma_{21}^\kappa}{\partial t} \approx 0$ and $\hbar \frac{\partial \rho_{22}^\kappa}{\partial t} \approx 0$ [5]. To solve Eqs.(5) and (6) in the linear region $\rho_{11} \gg \rho_{22}$ (see support information for the detailed discussion), we obtain

$$\sigma_{21}^{\kappa} = \frac{C_3^{\kappa'}C_2^{\kappa} - C_3^{\kappa}C_1^{\kappa'}}{C_1^{\kappa}C_1^{\kappa'} - C_2^{\kappa}C_2^{\kappa'}} \qquad (7)$$

and

$$\rho_{22}^{\kappa} = \frac{2\gamma_{21}}{\gamma_{22}}|\sigma_{21}^{\kappa}|^2 \qquad (8)$$

with $C_1^{\kappa} = \left[i\hbar(\omega - \omega_0) - \gamma_{21}\right] - iG_{\kappa}^{\kappa}$, $C_2^{\kappa} = -iG_{\kappa'}^{\kappa}$, $C_3^{\kappa} = i(\boldsymbol{\mu}_{21}^{\kappa} \cdot \boldsymbol{E}' + \boldsymbol{m}_{21}^{\kappa} \cdot \boldsymbol{B}')|_{r=r_{\kappa}}$, $G_{\kappa'}^{\kappa} = -\boldsymbol{E}_d^{\kappa'}(\boldsymbol{r}_{\kappa}) \cdot \boldsymbol{\mu}_{21}^{\kappa}$ and $G_{\kappa}^{\kappa} = -\boldsymbol{E}_d^{\kappa}(\boldsymbol{r}_{\kappa}) \cdot \boldsymbol{\mu}_{21}^{\kappa}$. Thus, the absorption rate of the molecule $\kappa$ can be expressed as

$$Q_{mol}^{\kappa} = \omega_0 \rho_{22}^{\kappa} \gamma_{22} = 2\omega_0 \gamma_{21} \frac{\left|C_1^{\kappa'}(\boldsymbol{\mu}_{21}^{\kappa} \cdot \boldsymbol{E}' + \boldsymbol{m}_{21}^{\kappa} \cdot \boldsymbol{B}')|_{r=r_{\kappa}} - C_2^{\kappa}(\boldsymbol{\mu}_{21}^{\kappa'} \cdot \boldsymbol{E}' + \boldsymbol{m}_{21}^{\kappa'} \cdot \boldsymbol{B}')|_{r=r_{\kappa'}}\right|^2}{|C_1^{\kappa}C_1^{\kappa'} - C_2^{\kappa}C_2^{\kappa'}|^2} \qquad (9)$$

After some algebra, we can obtain

$$CD_{mol} = \sum_{\kappa=1}^{2} CD_{mol}^{\kappa} \qquad (10)$$

with

$$CD_{mol}^{\kappa} = \frac{8\sqrt{\varepsilon_r}\,\omega_0 \gamma_{21}|E_0|^2}{3c\,|C_1^{\kappa}C_1^{\kappa'} - C_2^{\kappa}C_2^{\kappa'}|^2}\,\text{Im}\left\{\left(C_1^{\kappa'}\boldsymbol{m}_{21}^{\kappa} - C_2^{\kappa}\boldsymbol{m}_{21}^{\kappa'}\right)\left[\hat{P}^{\dagger}(\boldsymbol{r}_{\kappa})(C_1^{\kappa'}\boldsymbol{\mu}_{21}^{\kappa})^* - \hat{P}^{\dagger}(\boldsymbol{r}_{\kappa'})(C_2^{\kappa}\boldsymbol{\mu}_{21}^{\kappa'})^*\right]\right\}.(11)$$

Here $\hat{P}(\boldsymbol{r}_{\kappa})$ is the field enhanced matrix in the position of the molecule $\kappa$, which describes strong changes of the optical electric field inside the molecule due to the presence of NPs and it is expressed as [39]

$$\hat{P} = \begin{bmatrix} p_{xx} & p_{xy} & p_{xz} \\ p_{yx} & p_{yy} & p_{yz} \\ p_{zx} & p_{zy} & p_{zz} \end{bmatrix}. \qquad (12)$$

The matrix elements ($p_{\beta\alpha}$) in Eq.(12) are determined by the relation: $p_{\beta\alpha} = (\vec{E}_{ext})_{\beta}/|\vec{E}_0^{\alpha}|$, $\vec{E}_0^{\alpha}$ is a "$\alpha$" directional linear polarized plane wave and $\vec{E}_{ext}$ represents the field outside the circumscribing spheres, which can be obtained by the T-matrix method (see support information). As can be seen from Eq.(11), the CD signal of the molecules can be strongly modulated by the scattering field of NPs, because the field enhanced matrix elements possess very large values in the plasmonic resonance case.

**B . The absorption rate of NPs and $CD_{NP}$**

The absorption rate from NPs can be expressed as

$$Q_{NP}(\omega) = \sum_i Q_i(\omega) = 2\omega \sum_i (\text{Im}\varepsilon_{NP}) \int |E_{i\,tot}|^2 dV \qquad (13)$$

in which $\varepsilon_{NP}$ is the permittivity of NPs and the integral is taken over nanoparticle volumes, $E_{i\,tot} = E'_i + E^1_{d\,in} + E^2_{d\,in}$ denotes the total field inside the $ith$ nanoparticle ( $i = 1, 2 ......N$ ), $E'_i = \hat{K}E_0$ being the internal field generated by the plane wave in the absence of molecules. The position-dependent matrix determines the field inside NPs, which can be expressed as [39]

$$\hat{K} = \begin{bmatrix} k_{xx} & k_{xy} & k_{xz} \\ k_{yx} & k_{yy} & k_{yz} \\ k_{zx} & k_{zy} & k_{zz} \end{bmatrix} \qquad (14)$$

The matrix elements ( $k_{\beta\alpha}$ ) in Eq.(14) are determined from the relation: $k_{\beta\alpha} = (\vec{E}^\alpha_{\text{int}})_\beta / |\vec{E}^\alpha_0|$ , where $\vec{E}^\alpha_{\text{int}}$ is the induced internal field of the wave from NPs in the cluster, which can be obtained by the T-matrix method (see supplement information). $E^1_{d\,in}$ and $E^2_{d\,in}$ are the fields generated by two molecules, respectively. The $CD_{NP}$ mainly comes from three terms: $CD_{NP} = CD_{NP-DF} + CD_{NP-DD} + CD_{NP-FF}$ , they can be expressed as:

$$CD_{NP-DF} = 4\omega(\text{Im}\,\varepsilon_{NP})\,\text{Re}\left\langle \int \left[ \left(E'^{*}_i \cdot E_{d\,in}\right)_+ - \left(E'^{*}_i \cdot E_{d\,in}\right)_- \right] dV \right\rangle_\Omega, \qquad (15)$$

$$CD_{NP-DD} = 2\omega(\text{Im}\,\varepsilon_{NP})\,\text{Re}\left\langle \int \left[ |E_{d\,in}|^2_+ - |E_{d\,in}|^2_- \right] dV \right\rangle_\Omega, \qquad (16)$$

$$CD_{NP-FF} = 2\omega(\text{Im}\,\varepsilon_{NP})\,\text{Re}\left\langle \int \left[ |E'_i|^2_+ - |E'_i|^2_- \right] dV \right\rangle_\Omega, \qquad (17)$$

with $\boldsymbol{E}_{d\,in} = \boldsymbol{E}_{d\,in}^{1} + \boldsymbol{E}_{d\,in}^{2}$. The terms $CD_{NP-DF}$, $CD_{NP-DD}$ and $CD_{NP-FF}$ all originate in the chiral currents inside NPs. The difference is that the first two terms are induced by electric and magnetic dipoles of chiral molecules, while $CD_{NP-FF}$ are generated by the multipole interactions between different NPs caused only by the incidence wave. $CD_{NP-DF}$ and $CD_{NP-DD}$ can be calculated following the procedure given in Ref.[5] except more complex algebra are needed, and the results are written as:

$$CD_{NP-DF} = \frac{8\omega\sqrt{\varepsilon_r}|E_0|^2}{3c} \sum_{i}\sum_{\kappa=1}^{2} \mathrm{Im}\,\varepsilon_{NP}\,\mathrm{Re}\int\left\{\frac{\boldsymbol{m}_{21}^{\kappa'}\cdot\left\{K^{\dagger}(\boldsymbol{r})\left(C_2^{\kappa}\boldsymbol{E}_{d\,i}\big|_{\mu_{12}^{\kappa}} - C_1^{\kappa}\boldsymbol{E}_{d\,i}\big|_{\mu_{12}^{\kappa'}}\right)\right\}}{C_1^{\kappa}C_1^{\kappa'} - C_2^{\kappa}C_2^{\kappa'}}\right\}dV \quad (18)$$

and

$$CD_{NP-DD} = 2\omega(\mathrm{Im}\,\varepsilon_{NP})\sum_{i}\sum_{\kappa=1,2}\left\{D_A \int \left|\boldsymbol{E}_{d\,i}\big|_{\mu_{12}^{\kappa}}\right|^2 dV + D_B \int \left[\boldsymbol{E}_{d\,i}\big|_{\mu_{12}^{\kappa}}\right]^{*}\cdot\left[\boldsymbol{E}_{d\,i}\big|_{\mu_{12}^{\kappa'}}\right]dV\right\} \quad (19)$$

with

$$D_A = \left\langle |\sigma_{21}^{\kappa}|^2 \right\rangle_{\Omega+} - \left\langle |\sigma_{21}^{\kappa}|^2 \right\rangle_{\Omega-} = \frac{1}{2\gamma_{21}\omega_0}\left[\left\langle Q_{mol}^{\kappa}\right\rangle_{+} - \left\langle Q_{mol}^{\kappa}\right\rangle_{-}\right] \quad (20)$$

and

$$\begin{aligned}
D_B &= \left\langle \sigma_{21}^{\kappa *}\sigma_{21}^{\kappa'}\right\rangle_{\Omega+} - \left\langle \sigma_{21}^{\kappa *}\sigma_{21}^{\kappa'}\right\rangle_{\Omega-} = \frac{2\sqrt{\varepsilon_r}|E_0|^2}{3c\left|C_1^{\kappa}C_1^{\kappa'} - C_2^{\kappa}C_2^{\kappa'}\right|^2}\Big\{ \\
&\quad iC_2^{\kappa}C_2^{\kappa'}\left[-\boldsymbol{m}_{21}^{\kappa}\cdot(P^{\dagger}(\boldsymbol{r}_{\kappa'})\boldsymbol{\mu}_{12}^{\kappa'}) + \boldsymbol{m}_{12}^{\kappa'}\cdot(P^{T}(\boldsymbol{r}_{\kappa})\boldsymbol{\mu}_{21}^{\kappa})\right] \\
&\quad -2C_2^{\kappa *}C_1^{\kappa}\,\mathrm{Im}\left[\boldsymbol{m}_{21}^{\kappa'}\cdot(P^{\dagger}(\boldsymbol{r}_{\kappa'})\boldsymbol{\mu}_{12}^{\kappa'})\right] - 2C_1^{\kappa'*}C_2^{\kappa'}\,\mathrm{Im}\left[\boldsymbol{m}_{21}^{\kappa}\cdot(P^{\dagger}(\boldsymbol{r}_{\kappa})\boldsymbol{\mu}_{12}^{\kappa})\right] \\
&\quad +iC_1^{\kappa'*}C_1^{\kappa}\left[-\boldsymbol{m}_{21}^{\kappa'}\cdot(P^{\dagger}(\boldsymbol{r}_{\kappa})\boldsymbol{\mu}_{12}^{\kappa}) + \boldsymbol{m}_{12}^{\kappa}\cdot(P^{T}(\boldsymbol{r}_{\kappa'})\boldsymbol{\mu}_{21}^{\kappa'})\right]\Big\}
\end{aligned} \quad (21)$$

where $\boldsymbol{E}_{d\,i}\big|_{\mu_{12}^{\kappa}}$ is the electric field inside the $i^{th}$ nanoparticle induced by a dipole with a momentum of $\mu_{12}^{\kappa}$, which can also be calculated by the T-matrix method (see supplement information). When the two-level molecule is located in the hot spot, radiation from its electric dipole is strongly enhanced, the interaction between the molecule and NPs become strong, and the values of $CD_{NP-DF}$ and $CD_{NP-DD}$ become large especially in the plasmonic resonant case. In addition, the plasmon-mediated interaction between two molecules can also be reflected from Eq.(19) because of the existence of cross terms $D_B \int \left[ \bm{E}_{di} \big|_{\mu_{12}^{\kappa}} \right]^* \cdot \left[ \bm{E}_{di} \big|_{\mu_{12}^{\kappa'}} \right] dV$.

The term $CD_{NP-FF}$ represents the CD induced by the structure in the absence of chiral molecules (structure chirality). This term is zero under the long wave approximation, beyond the quasi-static approximation it can be rewritten as:

$$CD_{NP-FF} = \langle C_{abs+} - C_{abs-} \rangle_\Omega I_{inc} \qquad (22)$$

where $I_{inc} = 2\varepsilon_{vac} c \sqrt{\varepsilon_r} |E_0|^2$ is the intensity of the incident circularly polarized plane wave, and $C_{abs\pm}$ are the absorption cross sections of the NPs irradiated by the left- and right-handed polarized lights. The detailed calculation process for $\langle C_{abs\pm} \rangle_\Omega$ by the T-matrix method is given in the supplement information. Here the CD has been defined as the difference between absorption of left- and right-handed polarized lights[5, 14]. This is because the sizes of particles are far less than the wavelength of the incident light. In this case defining CD by absorption or extinction makes no difference. Based on the above equations, the CD of hybrid systems consisting of two molecules and nanoparticle clusters can be obtained through numerical calculations.

### III. NUMERICAL RESULTS AND DISCUSSION.

**A. Molecule-induced plasmonic CD in molecule-NP nanocomposites**

In this part, we consider molecule-induced plasmonic CD in the molecule-NP nanocomposites without the structure chirality. We first calculate CD spectra for a chiral molecule inserted into the hot spots of silver nanosphere cluster. The CD for such a system has been discussed very well in the previous works[5, 39], which allows us to validate our calculated method. The dashed lines in Fig. 2 represent the calculated CD signals two times as a function of wavelengths for the hybrid system consisting of a molecule and a linearly arranged trimer with an interparticle distance of 1.0 nm. The radii of silver spheres are taken to be 10nm. The parameters of the molecular dipole are taken according to Ref. [1, 5]: $\mu_{12} = |e| r_{12}$ and $\vec{\mu}_{12} \cdot \vec{m}_{21} / \mu_{12} = i|e| r_0 \omega_0 r_{21} / 2$. In our calculations, we use $r_{12} = 2 \overset{\circ}{A}$, $r_0 = 0.05 \overset{\circ}{A}$ and $\gamma_{12} = 0.3 eV$. For the dielectric functions of Ag, the Johnson's data were adopted [40], the permittivity of water is taken to be $\varepsilon_0 = 1.8$. If the CD signals are in units of $M^{-1} cm^{-1}$, they have to be multiplied by $NA/(0.23 I_{inc})$ where $NA = 6.02 \times 10^{23}$ is Avogadro's number [5, 14, 41]. Figure 2 (a), (b) and (c) correspond to $CD_{mol}$, $CD_{NP-DF}$ and $CD_{NP-DD}$, respectively, the total CD is given in Fig.2 (d). The resonance of molecules appears at $\lambda_0 = 300 nm$, which is away from the plasmon resonance for Ag spheres (about 400nm). Our calculated results are identical with those in Ref.[39].

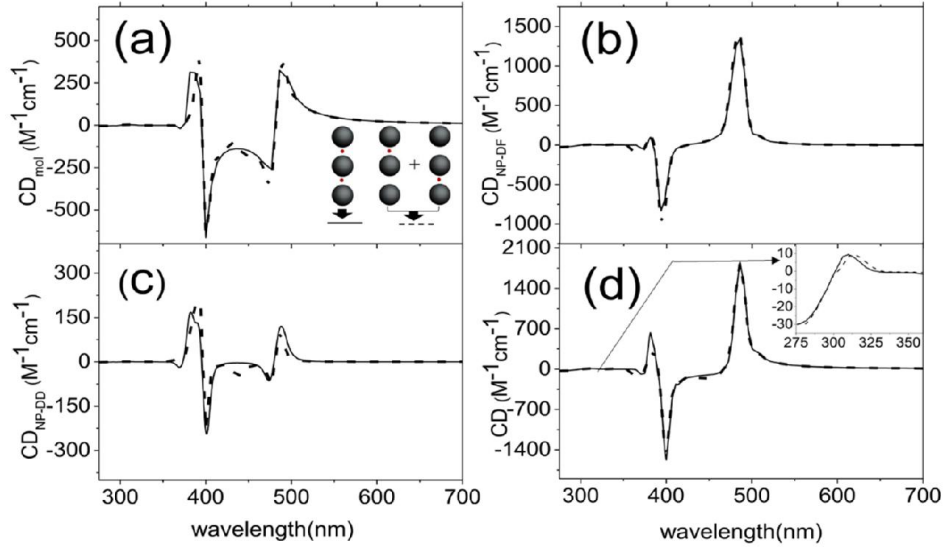

**Figure 2.** (Color online) Calculated CD signals as a function of wavelengths for molecule-NP complexes with three spheres as shown in (a). The solid lines describe the case when two chiral molecules are inserted at the same time into two hotspots of the linearly arranged silver trimer, the dashed lines correspond to the CD signals two times for the case when only one chiral molecule is put in the hotspot of the trimer. The radii of silver spheres are 10nm and separations between them are 1nm. Two molecules have dipoles $\mu_z||z$, and the resonance of molecules at $\lambda_0 = 300nm$. (a), (b), (c) and (d) correspond to $CD_{mol}$, $CD_{NP-DF}$, $CD_{NP-DD}$ and total CD, respectively. Inset in (d) exhibits CD signal around 300nm in different scales.

The solid lines in Fig. 2 display the corresponding results when two chiral molecules are inserted at the same time into two hot spots of the linearly arranged trimer. Comparing them with the dashed lines, we find that their values and trends are basically identical. Herein, the total CD of the system with two chiral molecules is the sum of the CD signal for each molecule inserted separately into the hot spot of the nanosphere trimer. If we take $\lambda_0 = 400nm$, which is near the plasmon resonance wavelength, the similar phenomenon appears. Figure 3 describes the corresponding results when separations between spheres are taken to be 2nm. The consistency between the solid lines and dashed lines is observed again.

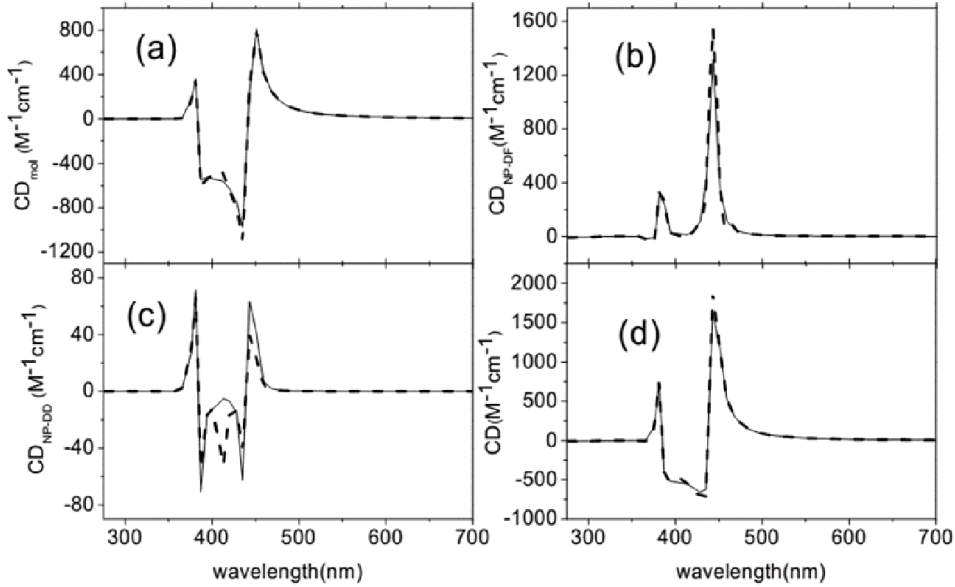

**Figure 3.** Calculated CD signals as a function of wavelengths for molecule-NP complexes with three spheres. The solid lines describe the case when two chiral molecules are inserted at the same time into two hotspots of the linearly arranged silver trimer, the dashed lines correspond to the CD signals two times for the case when only one chiral molecule is put in the hotspot of the trimer. (a), (b), (c) and (d) correspond to $CD_{mol}$, $CD_{NP-DF}$, $CD_{NP-DD}$ and total CD, respectively. The resonance of molecules is at $\lambda_0 = 400$nm and separations between spheres are 2nm. The other parameters are identical with those in Fig.2.

However, the situation becomes different when the interparticle distance decreases at $\lambda_0 = 400nm$. Figure 4 displays the corresponding results for the case with the interparticle distance of 1.0 nm as $\lambda_0 = 400nm$. The other parameters are identical with those in Fig. 3. It is seen clearly that they are not only different in amplitude for CD spectra but the positions of resonance peaks also shift, that is to say, the total CD of the system with two chiral molecules is not the sum of each single molecule signal when two chiral molecules are inserted respectively into the hot spot of nanosphere trimer. This is because the interaction between molecules and NPs becomes strong with the decrease of the interparticle distance. The strong interaction between molecules and NPs results in an indirect coupling through surface plasmonic excitations. For such a case, we have to consider the contribution of all molecules at the same time in calculating CD spectra of molecule-NP systems.

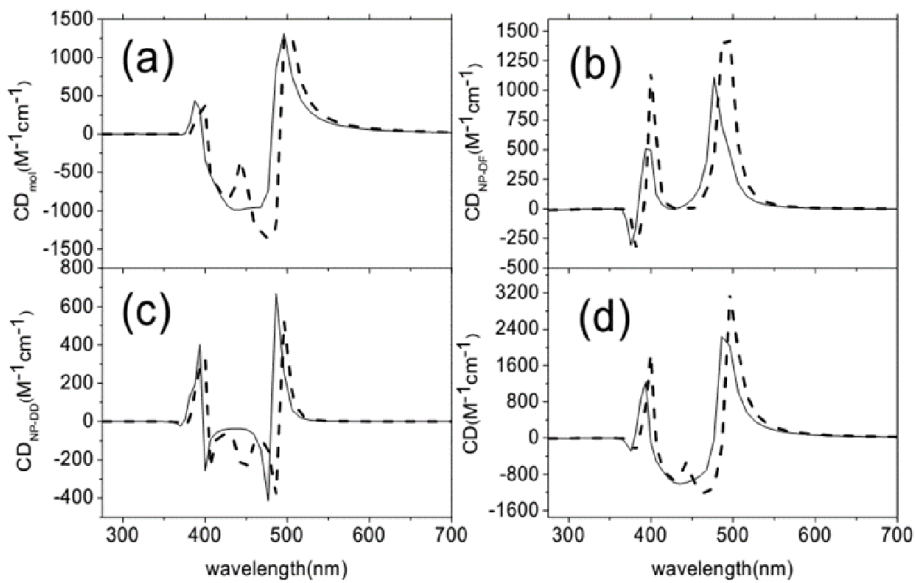

**Figure 4.** Calculated CD signals as a function of wavelengths for molecule-NP complexes with three spheres. The solid lines describe the case when two chiral molecules are inserted at the same time into two hotspots of the linearly arranged silver trimer, the dashed lines correspond to the CD signals two times for the case when only one chiral molecule is put in the hotspot of the trimer. (a), (b), (c) and (d) correspond to $CD_{mol}$, $CD_{NP-DF}$, $CD_{NP-DD}$ and total CD,

respectively. The resonance of molecules is at $\lambda_0 = 400$nm and separations between spheres are 1nm. The other parameters are identical with those in Figure.2.

**B. Plasmonic CD due to the intrinsic structure chirality**

In fact, the spatial arrangements of plasmonic NPs in nanocomposite can be inherently chiral [41, 42]. A lot of discussions have been done about the optical response from these structures. In order to prove the correctness of our method, we have performed calculations on the chirality from the chiral structures described in Refs.[42, 43], and obtained the same results with these references. Now we consider a dimer consisting of two spheroids in the absence of chiral molecules as shown in insets of Fig.5. If two spheroids possess the same axis orientation, the dimer is no-chiral. However, it can become the chiral structure through rotating the axis orientation of one spheroid. Figure 5 (a) shows calculated $CD_{NP-FF}$ as a function of wavelengths for a dimer of two silver spheroids with long axis a=10nm and two short axes b=9nm. Here, the separation between two spheroids is 2nm. The solid line, dashed line and dotted line correspond to angles of long axis orientation of one spheroid $\alpha_P = 0.01^0$, $0.02^0$ and $0.05^0$, respectively. With the increase of $\alpha_P$, the structure chirality becomes strong and $CD_{NP-FF}$ increases rapidly. When $\alpha_P = 0.05^0$, the value of $CD_{NP-FF}$ can be as high as $1.5 \times 10^6$, it is three-order larger than maxium of molecule-induced plasmon CD as shown in Fig. 2, 3 and 4.

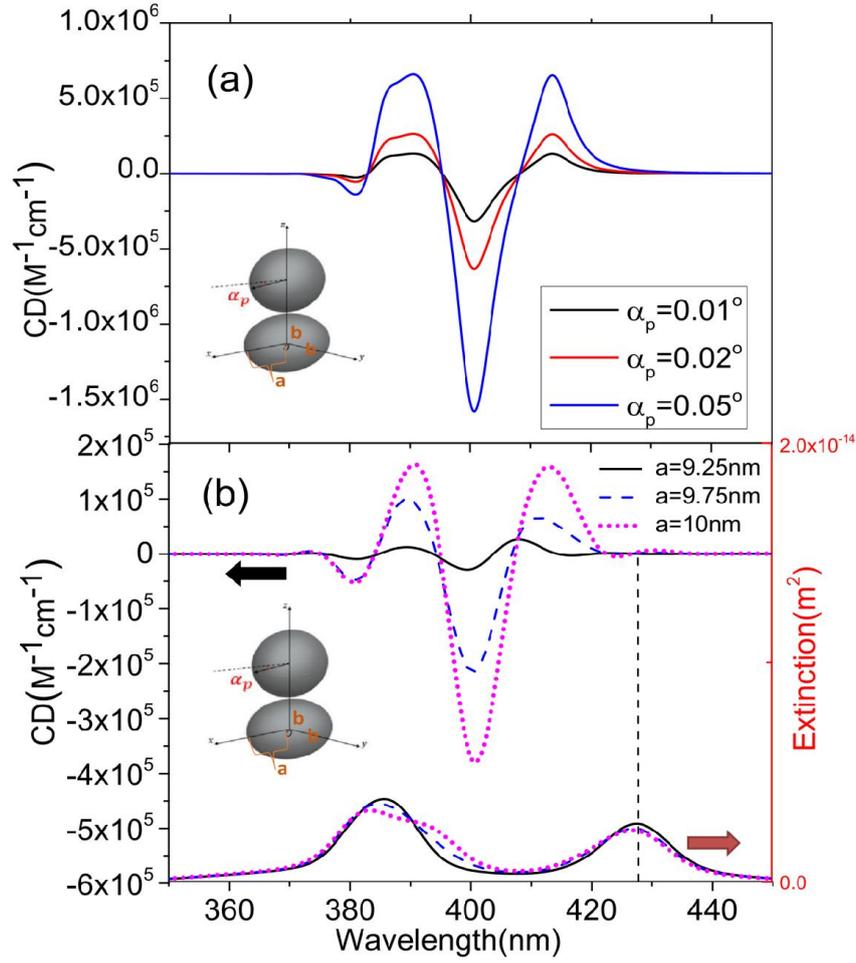

**Figure 5.** (Color online) (a) Calculated $CD_{NP-FF}$ as a function of wavelengths for a dimer of two silver spheroids with long axis a=10nm and two short axes b=9nm for various angles ($\alpha_p$) of long axis orientation of one spheroid. (b) The corresponding $CD_{NP-FF}$ for various lengths of long axes of spheroids with $\alpha_p = 0.01°$. Calculated extinctions for the corresponding dimer are shown. The separation between spheroids is 2nm.

The chirality of such a dimer is not only related to $\alpha_P$, it also depends on aspect ratio of the spheroid. The corresponding $CD_{NP-FF}$ for various ratios with $\alpha_p = 0.01°$ is plotted in Fig. 5 (b).

We can see clearly that the value of $CD_{NP-FF}$ increases largely with the length of the long axis of spheroids when the short axes are fixed. Comparing the spectra of $CD_{NP-FF}$ with the corresponding extinctions as shown in Fig. 5(b), we find that strong structure-induced CD only appears at the wavelength range below the plasmon coupling resonance region. At the plasmon coupling resonance region (around the wavelength 427nm), the value of $CD_{NP-FF}$ is very small in any case to change $\alpha_P$ and the ratio of long axis to short axis of the spheroid. This can be seen more clearly from Fig. 6.

Figure 6(b) shows calculated $CD_{FF}$ and extinction as a function of wavelengths for a linearly arranged trimer of three Au spheroids with long axis a=17.5nm and two short axes b=17nm. When the axis orientation of the middle spheroid is fixed, rotating other two spheroids as shown in Fig. 6(a), the trimer becomes a chiral structure. The black solid line and dashed line in Fig. 6(a) represent the CD at $\alpha_p = 0.01°$ as the gaps between nearest neighbor spheroids are taken to be 1nm and 2nm, respectively. The corresponding extinctions are described by the red solid line and dashed line. Here the parameters of Au are also taken according to Ref.[40]. With the decrease of the gap, the plasmon coupling resonance peak shifts to longer wavelength (redshift), while the CD curves do not change a lot in their peaks or dips. The point A (528.03nm) and B (653.09nm) in the extinction curve correspond to the single scattering and coupling plasmon resonance peaks, respectively. The CD in the coupling resonance region is small and changes monotonously as a function of wavelength. In order to disclose the physical origin of the phenomenon, in panels A and B of Figure 6 we plot the electric filed intensities in xz plane at A and B points as marked in Fig. 6(b), which is excited by the left-handed circularly polarized light. The fields between the scattering object and the smallest circumscribing sphere are calculated by the method of near-field T-matrix introduced in Ref.[44]. In contrast to the field distribution at the point A, the electric field distribution at the point B mainly focuses in the gap region. In keeping the size of gap unchanged, the effect of the axis orientation of spheroids on such a field distribution is small, and makes it harder to generate chiral currents or charges due to rotations. This is in contrast to another kind of chiral nanostructure caused by dislocation of nanoparticles.

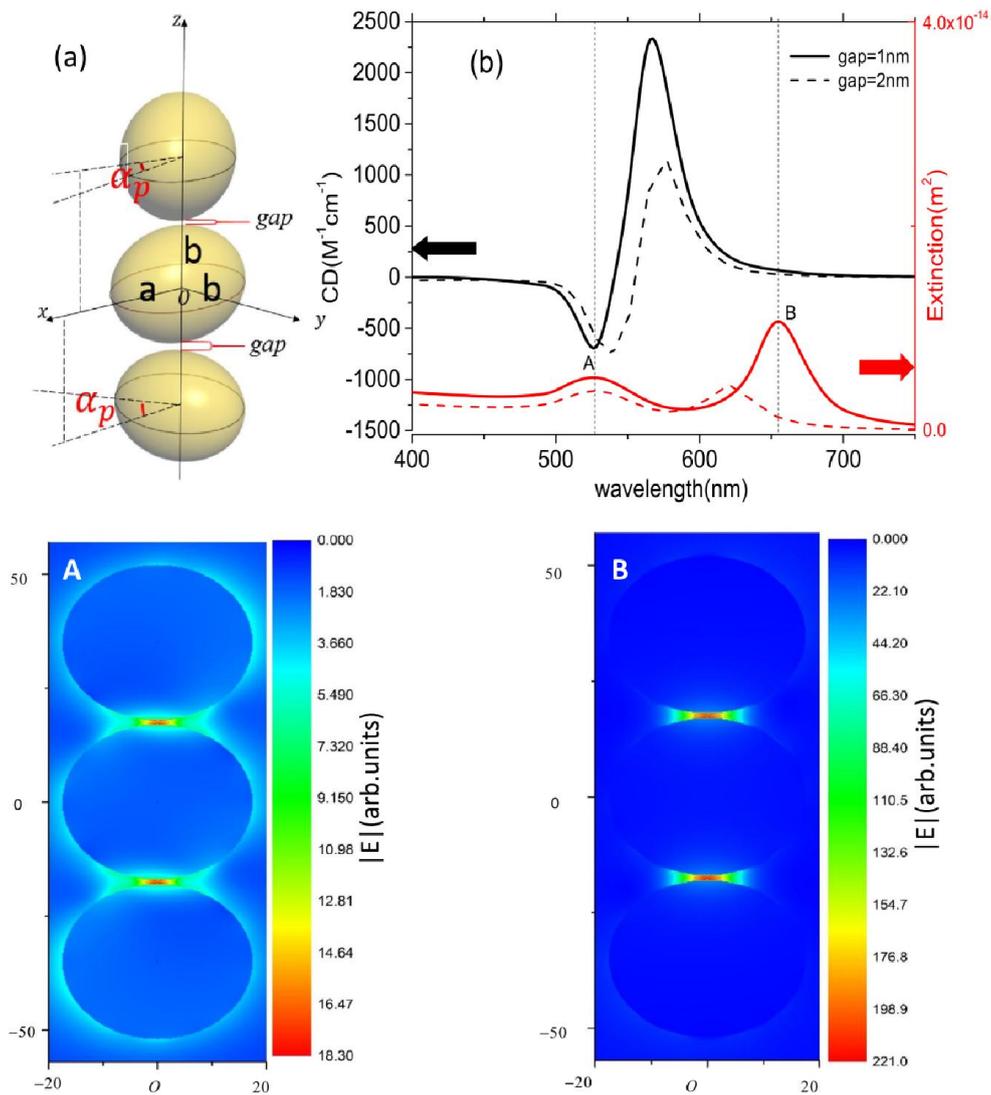

**Figure 6.** (Color online) (a) Calculated $CD_{NP-FF}$ and extinction as a function of wavelengths for the system of three Au spheroids with long axis a=17.5nm, two short axes

b=17nm and $\alpha_p = 0.01°$ for various gaps as shown in (b). Panels A and B represent spatial profiles of electric field amplitudes in xz plane at A (528.03nm) and B (653.09nm) points as marked in (a), which is excited by the left-handed circularly polarized light.

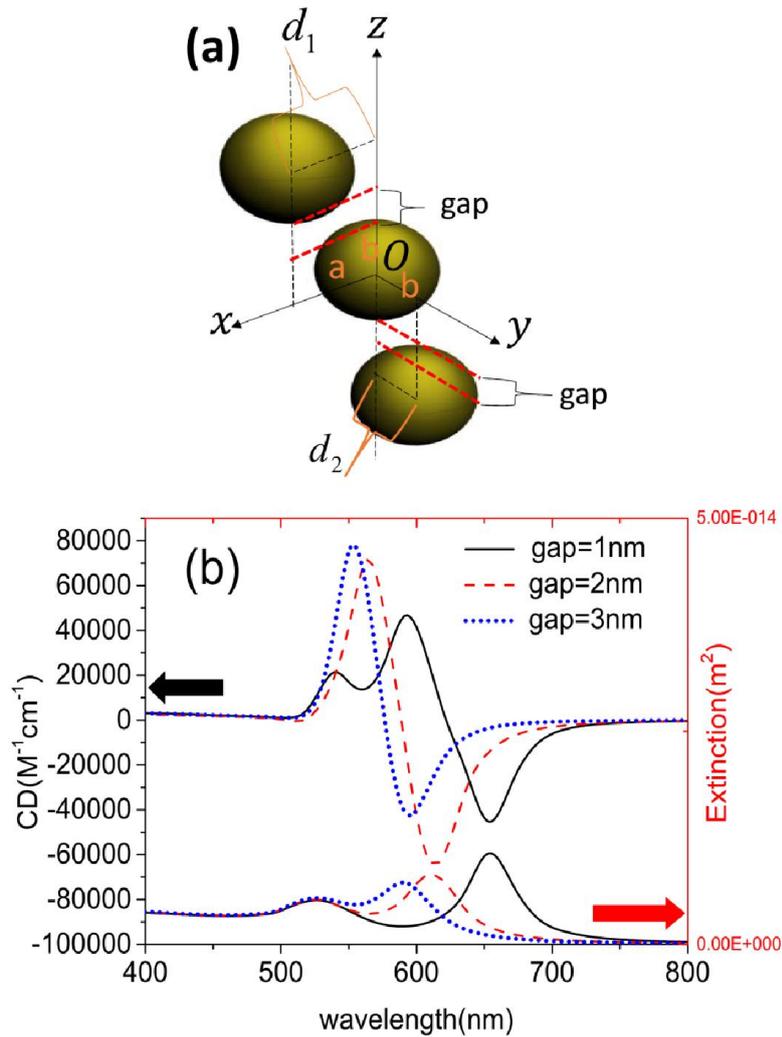

**Figure 7.** (Color online) (a) Schematics of spheroid trimer chiral structure caused by dislocation of nanoparticles. (b) Calculated $CD_{NP-FF}$ and extinction as a function of wavelengths

for the Au spheroid trimer chiral structure caused by dislocation of nanoparticles with a=17.5nm, b=17nm and $d_1=d_2=1nm$. The other parameters are identical with those in Figure.5.

For a linearly arranged trimer consisting of three spheroids, the structure chirality can also be produced by changing positions of spheroids instead of rotating the axis orientations of spheroids. If displacements of spheroids occur in a plane, the structure is still non-chiral. If displacements of spheroids occur as shown in Fig.7(a), that is, the position of the middle spheroid is fixed, one of the other two spheroids moves along the x axis with a distance $d_1$ and another along the y axis with a distance $d_2$, the trimer becomes a chiral structure. Figure 7(b) shows calculated $CD_{FF}$ and extinction as a function of wavelengths for such a system with a=17.5nm, b=17nm and $d_1=d_2=1nm$ for various gaps. The solid line, dashed line and dotted line correspond to the case when the gaps are taken to be 1nm, 2nm and 3nm, respectively. Comparing them with the corresponding extinction, we find that the coupling resonance peak from the extinction and the peak of $CD_{FF}$ have a close correspondence. That is to say, large value of $CD_{FF}$ appears at the coupling resonance region. This is in contrast to the case described in Fig.5 and 6.

**C. Competition of molecule-induced plasmon chirality and structure chirality**

If the chiral molecules are put in the hot spots of the chiral structures described in Fig. 6 and 7, the CD spectra of the system include two contributions, the structure chirality ($CD_{FF}$) and the molecule-induced plasmon chirality ( $CD_{mol} + CD_{NP-DD} + CD_{NP-DF}$ ). Figure 8(a) presents the comparison between them for the chiral structure described in Fig. 7.

The dashed line in Fig. 8(a) represents the structure chirality, the dotted line corresponds to the molecule-induced plasmon chirality, and the total CD is described by the solid line. It can be seen that the molecule-induced plasmon chirality is completely suppressed by the structure chirality. That is to say, the total CD can not exhibit the information of molecule chirality. If we want to detect chirality of chiral molecules by using surface plasmon resonances, we have to overcome the structure chirality, for example, controlling all nanoparticles in the same plane. In fact, a practical measurement will always need to average over many such structure, and therefore it could happen that the structure chirality will average to zero. At such a case, providing that there is some non-random molecular alignment with respect to the structure, the total CD can still exhibit the information of molecule chirality.

However, for the cases of chiral structures described in Fig. 5 and 6, we can still use them to detect molecular chirility as shown in Fig. 8(b) even the structure chirality does not average to zero. Figure 8(b) displays the comparison between the structure chirality and the molecule-induced plasmon chirality for the chiral structure described in Fig. 6. The solid line, dashed line and dotted line in Fig. 8(b) represent the total CD, the structure chirality and the molecule-induced plasmon chirality, respectively. Because the structure chirality is very weak and affected small by chiral structures around the plasmon coupling resonant peak (650nm), the total CD around 650nm can exhibit information of the molecule-induced plasmon chirality. That is to say, the chirality of chiral molecule can be detected from CD spectra in such a case.

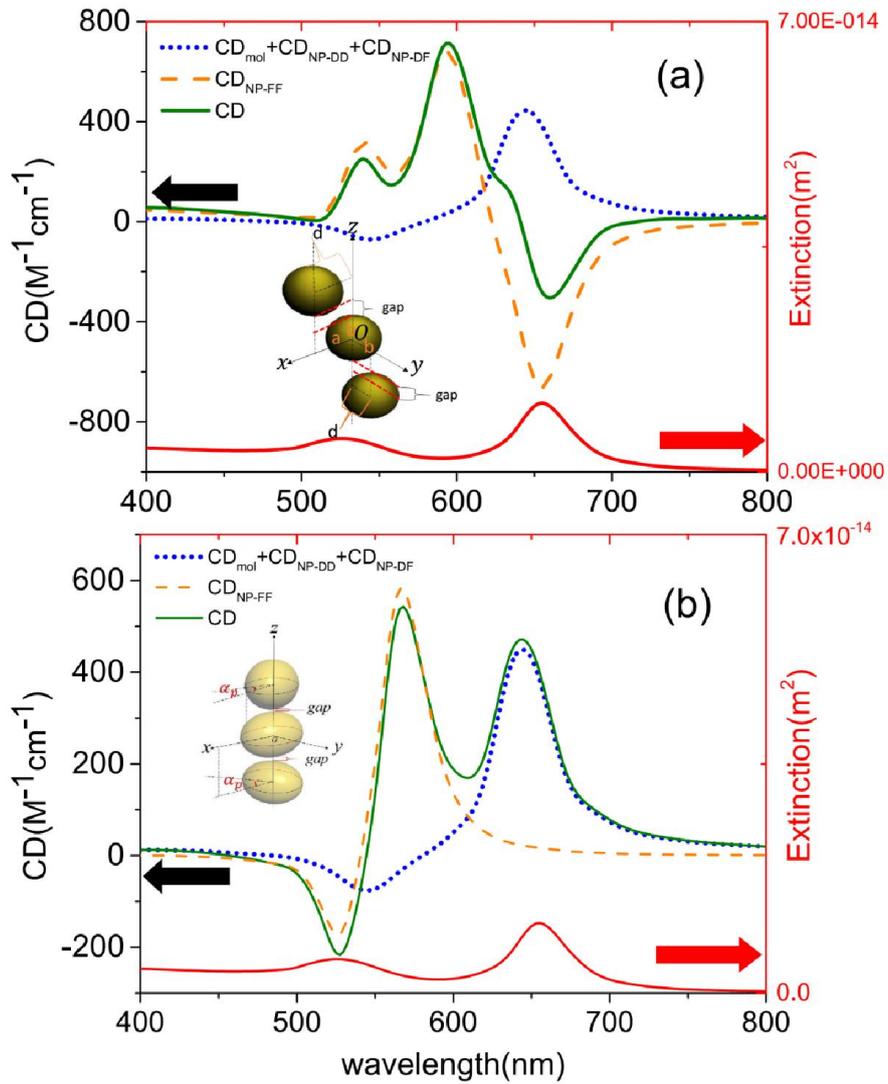

**Figure 8** (Color online) (a) Calculated CD and extinction as a function of wavelengths for the Au spheroid trimer chiral structure caused by dislocation of nanoparticles with a=17.5nm, b=17nm and $d_1 = d_2 = 1nm$. One molecule is placed at x=0.0, y=0.0 and z=17.5nm, the other is at x=0.0, y=0.0 and z=-17.5nm. (b) Calculated $CD$ and extinction as a function of wavelengths for the Au spheroid trimer chiral structure caused by rotating nanoparticles with a=17.5nm, b=17nm, $\alpha_p = 0.0025°$ and gap = 1nm. The solid line, dashed line and dotted line represent the total CD,

the structure chirality ($CD_{NP-FF}$) and the molecule-induced plasmon chirality ($CD_{mol} + CD_{NP-DD} + CD_{NP-DF}$), respectively, the dot-dashed line is extinction. Here $\lambda_0 = 300$nm and $\mu_z \| z$.

**Summary and discussion.** We have presented a T-matrix method to calculate CD signals of the hybrid system consisting of two molecules and a cluster with arbitrarily disposed nanoparticles. The CD spectra for the chiral molecules inserted into clusters with various metallic NPs have been calculated. Our calculated results have shown that the total CD of the system with two chiral molecules is not sum for two cases when two chiral molecules inserted respectively into the hot spots of nanoparticle clusters as the distances among nanoparticles are small, although the relationship is established at large interparticle distances. The structure chiralities from two kinds of chiral structure, caused by rotating spheroids or dislocation, have also been investigated. The structure chirality generally plays a leading role in the CD spectrum for a hybrid system consisting of molecules and chiral structures. Therefore, how to reduce the contribution from the structure chirality is a key issue to realize ultrasensitive probe of chiral information of molecules by plasmon-based nanotechnology. We find that, for some special chiral structures, the contribution by the structure chirality to the CD signal can be very small at plasmon coupling resonance regions thus the molecule-induced plasmon chirality makes predominant role in determineing the probed CD signal. We believe that our findings can provide an important reference for ultrasensitive detection of molecular chirality, a key aspect for various bioscience and biomedicine applications.

**AUTHOR INFORMATION**

**Corresponding Author**

\* E-mail: zhangxd@bit.edu.cn

**Notes**

The authors declare no competingfinancial interest.

**Acknowledgment**

We wish to thank Hui Zhang for useful discussions. This work was supported by the National Natural Science Foundation of China (Grant No. 11274042) and the National Key Basic Research Special Foundation of China under Grant 2013CB632704.

**Supporting Information**

Details of analytical derivations for theory and method are given. This material is available free of charge via the Internet at http://pubs.acs.org.

**The Table of Contents graphic:**

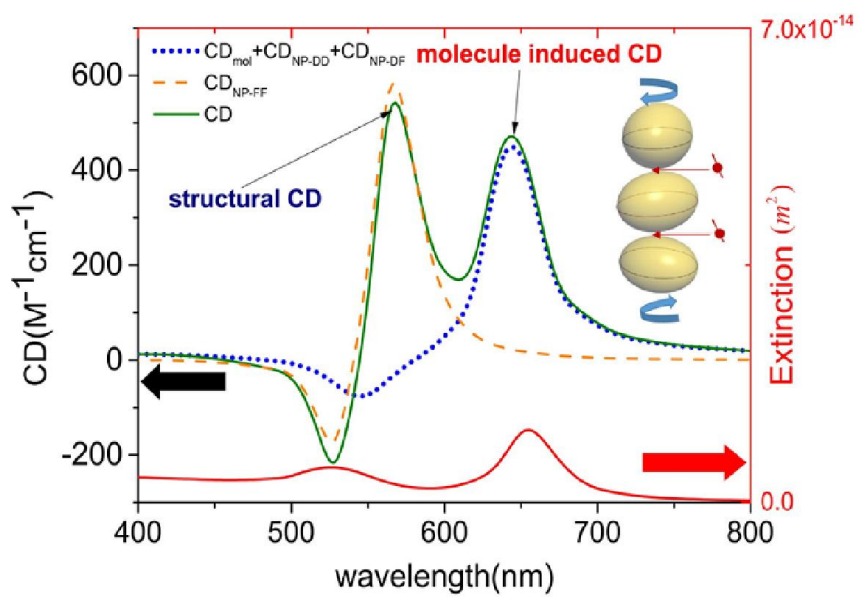